\documentclass[12pt]{article}
\usepackage{amsmath}

\parskip=\medskipamount	
\newcommand{\bea}{\begin{eqnarray}}
\newcommand{\eea}{\end{eqnarray}}
\newcommand{\bth}{{\boldsymbol\theta}}
\newcommand{\bkp}{{\boldsymbol\kappa}}
\newcommand{\bpi}{{\boldsymbol\pi}}

\begin{document} 

\thispagestyle{empty}

\begin{flushright}
\begin{tabular}{l}
CTP-MIT-3149 \\ 
BUHEP-01-11\\ 
RU-01-10-B 
\end{tabular} 
\end{flushright}

\vspace{.5cm}

\begin{center}
{\LARGE {\bf Testing Non-commutative QED, \\[1ex]
Constructing Non-commutative MHD.}}
\vspace{.8cm} \

{Z. Guralnik$^{1}$, R. Jackiw$^{1}$,
S.Y. Pi$^{2}$ and A.P. Polychronakos$^{3,\dagger}$} \\
\vspace{0.5cm}

{$^1$ {\it Center for Theoretical Physics }} \\ 
{{\it Massachusetts Institute of Technology}} \\
{{\it Cambridge MA, 02139, USA }} \\
\vspace*{0.3cm}
{$^2$ {\it Physics Department, Boston University}} \\
{{\it Boston MA 02215, USA}}\\
\vspace*{0.3cm}
{$^3$ {\it Physics Department, Rockefeller University}} \\  
{{\it New York NY, 10021, USA}}  \\

\end{center}

\vspace{.5cm}

\begin{abstract}
The effect of non-commutativity on electromagnetic waves
violates Lorentz invariance: in the presence of a 
background magnetic induction field $\bf b$,  the velocity for
propagation transverse to $\bf b$ differs from $c$,  
while propagation along $\bf b$
is unchanged. In principle, this allows a test by the 
Michelson-Morley interference method. 
We also study non-commutativity in another context, by constructing 
the theory describing a charged fluid in a strong magnetic field, 
which forces the fluid particles 
into their lowest Landau level, and renders the fluid dynamics
non-commutative, with a Moyal product determined by the background
magnetic field.
\end{abstract}

\vspace{1.0cm}
\noindent
\rule{6.5cm}{0.4pt}

\noindent {\footnotesize ${}^{\dagger}$~On leave of absence from
Theoretical Physics Department, Uppsala  University, Uppsala 751~08,
Sweden, and University of Ioannina, Ioannina 45110, Greece.}

\setcounter{page}{0}  
\newpage 
\setcounter{footnote}{0}



\section{Introduction}

The idea that spatial coordinates do not commutate \cite{1}
has a well-known realization in physics:  the quantized motion of
particles in a magnetic field, sufficiently strong so that 
projection on the lowest Landau level can be justified, is
described by non-commuting coordinates on the plane perpendicular
to the field \cite{2}.  Recently this phenomenon has played a
role in various quantum mechanical studies, involving both 
theoretical models \cite{3} and phenomenological 
applications \cite{4}.  At the same time generalizations to
quantum field theory have also been made, giving rise to various 
``non-commutative'' field theories, for example non-commutative
quantum electrodynamics.

In Section 2 of this paper we examine the effect of an external 
magnetic field on non-commutative photon dynamics, i.e. 
electrodynamics without charged particles, which nevertheless is
a non-linear theory owing to its non-commutativity.  We show that
the velocity of light depends on the direction of propagation 
relative to the external magnetic field,  thus allowing for a 
Michelson Morley-type test of non-commutativity,
which evidently violates special relativity (Lorentz invariance).

In Section 3 we study a magnetohydrodynamical (MHD) field theory in 
an intense magnetic field,  which effects a field theoretical analog
for the previously mentioned reduction to the lowest Landau level,
and results in non-commutative MHD,  in complete analogy to what 
happens to particles in a strong magnetic field \cite{2}.
Here the non-commutativity manifests itself in the charged 
fluid density not commuting with itself.  The form of the
non-commutativity depends on the nature of the density: if the fluid
is structureless, with quantum commutators deduced 
from Poisson brackets,  one particular expression is obtained.  
When the fluid is constructed from point particles, whose coordinates 
do not commute, then the density commutator involves a Moyal phase,
which reduces to the previous expression in a semi-classical limit.
Relevent formulas are expressed succinctly with the help of the 
``star'' product. 

\section{Testing non-commutative QED}
\label{testing}

The non-commutative generalization for the free Maxwell Langrange
density involves the ``star'' product of the non-commutative field
strength $\hat F_{\mu\nu}$, constructed from the potential 
$\hat A_{\mu}$,
\bea
\hat F_{\mu\nu} = \partial_{\mu}\hat A_{\nu} - 
\partial_{\nu}\hat A_{\mu} - 
i g (\hat A_{\mu} * \hat A_{\nu} - \hat A_{\nu} * \hat A_{\mu}) 
\label{eq1a} \\
\hat {\cal L} = -\frac{1}{4} \hat F_{\mu\nu} * \hat F^{\mu\nu},
\label{eq1b}
\eea
where the star product is defined by
\bea
(f*g)(x) = e^{\frac{i}{2} 
\theta^{\alpha\beta}\partial_{\alpha}\partial_{\beta}^{\prime}}
f(x)g(x^{\prime})|_{x^{\prime} = x} .
\label{eq2}
\eea
The non-linear terms in (\ref{eq1a}) enter with the coupling 
$g = \frac{e}{\hbar c}$.  To first order in 
$\theta^{\alpha\beta} = - \theta^{\beta\alpha}$, $\hat {\cal L}$
may be expressed in terms of the conventional Maxwell tensor
\bea
F_{\mu\nu} = \partial_{\mu}A_{\nu} - \partial_{\nu}A_{\mu},
\label{eq3}
\eea
with $\hat A_{\mu}$ related to $A_{\mu}$ by
\bea
\hat A_{\mu} = A_{\mu} - \frac{1}{2} \theta^{\alpha\beta}
A_{\alpha}(\partial_{\beta}A_{\mu} + F_{\beta\mu}) \\
\hat F_{\mu\nu} = F_{\mu\nu} + \theta^{\alpha\beta}
F_{\alpha\mu}F_{\beta\nu} - \theta^{\alpha\beta}A_{\alpha}
\partial_{\beta}F_{\mu\nu},
\eea
with $g$ absorbed in $\theta$.
It follows that apart from a total derivative term, which does not
affect the equations of motion \cite{5},
\bea
\hat {\cal L} = -\frac{1}{4} F^{\mu\nu}F_{\mu\nu} 
+\frac{1}{8}\theta^{\alpha\beta}F_{\alpha\beta}F_{\mu\nu}F^{\mu\nu}
-\frac{1}{2}\theta^{\alpha\beta}F_{\mu\alpha}F_{\nu\beta}F^{\mu\nu}
+{\cal O}(\theta^2).
\label{eq4}
\eea

Our strategy is to solve the equations of motion implied by 
(\ref{eq4}) (to first order in $\theta$)  and to exhibit 
how special relativity is violated. Henceforth we take 
$\theta^{\alpha\beta}$ to have only spatial components, 
$\theta^{0\alpha} =0$, $\theta^{ij} = \epsilon^{ijk}\theta^k$,
and work exclusively with the field strengths $F^{i0} = E^i$
and $F_{ij} = -\epsilon_{ijk}B^k$, rather than with the vector
potential.

The ``Maxwell'' equations that follow from (\ref{eq4}) 
are
\bea
\frac{1}{c}\frac{\partial}{\partial t} \bf B + 
\bf\nabla \times \bf E = 0 \label{eq5a} \\
\bf \nabla \cdot \bf B =0. \label{eq5b}
\eea
These of course are a consequence of (\ref{eq3}).
The other equations reflect the non-linear dynamics of 
(\ref{eq4}),  and can be written in terms of a 
displacement field $\bf D$ and magnetic field $\bf H$.
\bea
\frac{1}{c} \frac{\partial}{\partial t}{\bf D} - 
\bf\nabla \times \bf H = 0 \label{eq6a}\\
\bf \nabla \cdot \bf D = 0
\label{eq6b}
\eea
Constitutive relations follow from (\ref{eq4}).
\bea
{\bf D} = (1 - {\bth} \cdot {\bf B}) {\bf E} + 
({\bth} \cdot {\bf E}) {\bf  B} + ({\bf E} \cdot 
{\bf B}) {\bth} \label{eq7a} \\
{\bf H} = (1 - {\bth} \cdot {\bf B}) {\bf B} + 
\frac{1}{2}({\bf  E}^2 - {\bf  B}^2) {\bth} - 
({\bth} \cdot {\bf E}) {\bf E}
\label{eq7b}
\eea
We seek solutions to (\ref{eq5a}) -- (\ref{eq7b}) where
the electric field is a propagating plane wave.
\bea
{\bf E} = {\bf E}(\omega t - {\bf k} \cdot {\bf r})
\label{eq8}
\eea
Equation (\ref{eq5a}) implies that 
\bea
{\bf B} = {\bkp} \times {\bf E} + {\bf b},
\label{eq9}
\eea
where ${\bkp} = c{\bf k}/\omega$ and
${\bf b}$ is a time-independent background magnetic induction 
field,  which must be transverse according to (\ref{eq5b}).
However, we shall specialize by taking ${\bf b}$ to be constant. 

From (\ref{eq7a}) -- (\ref{eq9}) it follows that ${\bf D}$
and ${\bf H}$ are functions of $\omega t - {\bf k} \cdot {\bf r}$, 
so (\ref{eq6a}) implies that
\bea
{\bf D} = - {\bkp} \times {\bf H} + {\bf d}
\label{eq10}
\eea
where again ${\bf d}$ is a time-independent transverse background.  We
assume that no background field contributes to $\bf D$,
so $\bf d$ is chosen to cancel the constant contribution to 
$- {\bkp} \times {\bf H}$ coming from ${\bf b}$.

After $\bf D$ and $\bf H$  are expressed in terms of $\bf E$
using (\ref{eq7a}),(\ref{eq7b}) and (\ref{eq9}), 
equation (\ref{eq10}) becomes
\bea
(1- {\bth} \cdot {\bf b})E^i + 
\epsilon^{ijk}\kappa^jE_T^k ({\bf E} \cdot {\bth}) 
- E^i\epsilon^{jkl} \theta^j \kappa^k E_T^l + \beta^{ij} E^j \nonumber \\
= \kappa^2 E_T^i - 
\kappa^2 E_T^i \epsilon^{jkl} \theta^j \kappa^k E_T^l
-\frac{1}{2} \epsilon^{ijk}\kappa^j\theta^k  E^2(1-\kappa^2)
-\frac{\kappa^2}{2}\epsilon^{ijk}\kappa^j\theta^k E_L^2 \nonumber \\
+\epsilon^{ijk}\kappa^j E_T^k ({\bf E} \cdot {\bth})
+ \kappa^2
(\hat \kappa^j \hat \kappa^k \beta^{jk} 
- \beta^j_j - {\bth} \cdot {\bf  b})E_T^i + \kappa^2\beta^{ij}E_T^j
-\kappa^i \kappa^j \beta^{jk}E_T^k
\label{eq11}
\eea
where $\hat \bkp$ is the unit vector 
${\bkp} /|{\bkp} |$, and 
$\beta^{ij} = \theta^i b^j+\theta^j b^i$. The electric field
has been decomposed into transverse and longitudinal parts,
${\bf E} = {\bf E}_T + \hat\bkp E_L$, with 
$\hat {\bkp} \cdot {\bf  E}_T =0$.  Note that 
$\epsilon^{ijk} \kappa^j E_T^k ({\bf  E} \cdot {\bf  \theta})$ 
cancels from both sides of the equality.  By projecting
the above on $\hat \bkp$, we arrive at an expression for
the longitudinal component of $\bf  E$.
\bea
(1 - {\bth} \cdot {\bf  b} - 
\epsilon^{jkl} \theta^j \kappa^k E_T^l)E_L 
+ \hat \kappa^i \beta^{ij}E^j = 0
\label{eq12}
\eea
To lowest order in $\theta$, this give for $E_L$
\bea
E_L = -\hat \kappa^i \beta^{ij} E_T^j.
\label{eq13}
\eea
Re-inserting this in (\ref{eq11}) and keeping terms at most linear
in $\theta$ leaves
\bea
&&(1-\kappa^2)\left[(1-{\bth} \cdot {\bf  b}
-\epsilon^{jkl}\theta^j \kappa^k E_T^l)
E_T^i 
+ \beta^{ij}E_T^j - \hat \kappa^i \hat \kappa^j \beta^{jk}E_T^k
+\frac{1}{2}\epsilon^{ijk}\kappa^j\theta^k E_T^2\right] = 
\nonumber \\
&&\kappa^2(\hat \kappa^j \hat \kappa^k \beta^{jk} - \beta^j_j)E_T^i.
\label{eq14}
\eea

In the absence of non-commutativity (${\bth} =0$) the above
reduces to $(1-\kappa^2)E_T^i = 0$, which implies the usual dispersion
result $\kappa^2 =1$ or  $\omega^2 = c^2k^2$.  The same dispersion law
holds 
when there is no background field 
$\bf  b$.  However, when both $\bth$ and $\bf  b$ are non-vanishing,
$\kappa^2 =1$ is no longer a solution; rather we must take $1-\kappa^2$ to
be ${\cal O} (\theta)$.  Then to lowest, linear
order in $\theta$, (\ref{eq13}) becomes
\bea
(1 - \kappa^2) {\bf  E}_T = 
(\hat \kappa^j \hat \kappa^k \beta^{jk} 
- \beta^j_j){\bf  E}_T = -2 {\bth}_T \cdot {\bf  b}_T {\bf  E}_T.
\label{eq15}
\eea
Thus the solution demands a modified dispersion law:
\bea
\kappa^2 = \frac{c^2k^2}{\omega^2} = 
1 + 2{\bth}_T \cdot {\bf  b}_T
\label{eq16}
\eea
or
\bea
\omega = ck(1- {\bth}_T \cdot {\bf  b}_T)
\label{eq17}
\eea

The same result may be obtained more quickly and easily, if less
reliably, by linearizing the constitutive equations (\ref{eq7a}) and
(\ref{eq7b})
around the background magnetic induction field $\bf  b$.  Then
(\ref{eq7a}) and (\ref{eq7b}) read
\bea
D^i = \varepsilon^{ij}E^j \label{eq18a} \\
H^i = (\mu^{-1})^{ij}B^j
\label{eq18}
\eea
where the electric permitivity is given by 
\bea
\varepsilon^{ij} = \delta^{ij}(1-{\bth}\cdot {\bf  b}) + \beta^{ij}
\label{eq19}
\eea
 and the inverse magnetic permeability by
\bea
(\mu^{-1})^{ij} =\delta^{ij}(1 - {\bth} \cdot {\bf  b}) - \beta^{ij}
\label{eq20}
\eea
It is now posited  that all dynamical fields are functions of 
$\omega t - {\bf  k} \cdot {\bf  r}$, and that $\bf  E$, $\bf  B$,
$\bf  D$ and $\bf  H$ have no background field contributions.
With 
\bea
{\bf  B} = {\bkp} \times {\bf  E}
\label{eq21}
\eea
and
\bea
{\bf  D} = -{\bkp} \times {\bf  H},
\label{eq22}
\eea
it follows from (\ref{eq18a}) and (\ref{eq18}) that
\bea
D^i = -\epsilon^{ijk}\kappa^j(\mu^{-1})^{kl}\epsilon^{lmn}
\kappa^m (\varepsilon^{-1})^{nq}D^q.
\label{eq23}
\eea
To first order in $\theta$
\bea
(\varepsilon^{-1})^{ij} = \delta^{ij}(1 + {\bth}\cdot{\bf  b})
-\beta^{ij}
\label{eq24}
\eea
Inserting this and (\ref{eq20}) into (\ref{eq23}) gives
\bea
D^i = \kappa^2(1 + \hat\kappa^j\hat\kappa^k\beta^{jk} - \beta_j^j)D^i
\label{eq25}
\eea
whose solution is again (\ref{eq17}).

To recapitulate, we see that a plane electromagnetic wave does 
not see the non-commutativity 
if the background magnetic induction field $\bf  b$ vanishes,
or if the wave propagates in the direction of $\bf b$.
On the other hand, propagation transverse to $\bf  b$ is at a 
velocity that differs from $c$ by the factor 
$1-{\bth}_T \cdot {\bf  b}_T$. Note that both polarizations travel at the same (modified) velocity, so there is no Faraday-like rotation.
Let us also observe that the effective Lagrange density (\ref{eq4})
possesses two interaction terms proportional to $\theta$, 
with definite numerical constants.  Owing to the freedom
of rescaling $\theta$, only their ratio is significant.   It is
straightforward to verify that if the ratio is different 
from what is written in (\ref{eq4}), the two linear polarizations
travel at different velocities. Thus the non-commutative theory is 
unique in affecting the two polarizations equally, at least to 
${\cal O}(\theta)$ \cite{6}.

The change in velocity for motion relative to an external magnetic 
induction ${\bf b}$ allows searching for the effect with a 
Michelson-Morley experiment. In a conventional apparatus with
two legs of length $\ell_1$ and $\ell_2$ at right angles to each
other, a light beam of wavelength $\lambda$ is split in two, and
one ray travels along ${\bf b}$ (where there is no effect), while
the other, perpendicular to ${\bf b}$, feels the change of velocity
and interferes with the the first. After rotating the apparatus by 
$90^{\circ}$, the interference pattern will shift by 
$2(\ell_1 + \ell_2){\bth}_T \cdot {\bf b}_T/ \lambda$ fringes.
Taking light in the visible range, $\lambda \sim 10^{-5}$cm, 
a field strength $b \sim 1$ tesla, and using the current bound on
$\theta \le (10 TeV)^{-2}$ obtained in \cite{7}, one finds that
a length $\ell_1 + \ell_2 \ge 10^{18} cm \sim 1$ parsec
would be required for a shift of one fringe.  Galactic magnetic 
fields are neither that strong nor coherent over such large distances,
so another experimental setting needs to found to test for
non-commutativity.

Finally we note that there is close connection between our results on
photon propagation and the general analysis of Lorentz non-invariant
modifications to the standard model \cite{8}.

Added Note: A result identical to ours has been reported by 
R.-G. Cai, Phys.\ Lett.\ {\bf B} (in press), hep-th/0106047.

\section{Constructing non-commutative MHD}
\label{MHD}

\subsection{Particle non-commutativity in the lowest Landau level}

In order to describe the motion of a charged fluid in an intense
magnetic field, which effects a reduction to the field-theoretical 
analog of the lowest Landau level and results in a non-commutative
field theory,  we review the story for point particles on a plane,
with an external magnetic field $\bf b$ perpendicular to the plane
\cite{2}.
The equation for the 2-vector ${\bf r} = (x,y)$ is 
\bea
m \dot{v}^i = \frac{e}{c}\epsilon^{ij}v^jb+f^i({\bf r})
\label{eq31}
\eea
where $\bf v$ is the velocity $\dot{{\bf r}}$, and $\bf f$ represents
other forces, which we take to be derived from a potential $V$: 
${\bf f} = - {\bf\nabla}V$. The limit of large $b$ is equivalent to
small $m$. Setting the mass to zero in (\ref{eq31}) leaves a first
order equation.
\bea
\dot{r}^i = \frac{c}{eb}\epsilon^{ij}f^j({\bf r})
\label{eq32}
\eea
This may be obtained by taking Poisson brackets of $\bf r$ with 
the Hamiltonian
\bea
H_0 = V
\label{eq33}
\eea
provided the fundamental brackets describe non-commutative coordinates,
\bea
\{ r^i,r^j \} = \frac{c}{eb}\epsilon^{ij}
\label{eq34}
\eea
so that 
\bea
\dot{r}^i = \{ H_0, r^i \} = \{ r^j, r^i \}\partial_j V =
\frac{c}{eb}\epsilon^{ij}f^j({\bf r}).
\label{eq35}
\eea

The non-commutative algebra (\ref{eq34}) and the associated dynamics
can be derived in the following manner.
The Lagrangian for the equation of motion (\ref{eq31}) is
\bea
L = \frac{1}{2}m {v}^2 + \frac{e}{c}{\bf v} \cdot {\bf A} - V
\label{eq36}
\eea
where we choose the gauge ${\bf A} = (0,bx)$. Setting
$m$ to zero leaves
\bea
L_0 = \frac{eb}{c} x \dot{y} - V(x,y).
\label{eq37}
\eea
which is of the form $p\dot{q} - h(p,q)$, and one sees that
$(\frac{eb}{c}x,y)$ form a canonical pair.  This implies (\ref{eq34}),
and identifies $V$ as the Hamiltonian.  

Finally, we give a canonical derivation of non-commutativity in
the $m \rightarrow 0$ limit, starting with the Hamiltonian
\bea
H = \frac{ {\bf\pi}^2 }{2m} + V.
\label{eq38}
\eea
$H$ gives (\ref{eq31}) upon bracketing with $\bf r$, 
provided the following brackets hold;
\bea
&&\{ r^i, r^j \} = 0 \label{eq39a} \\
&&\{ r^i, \pi^j \} = \delta^{ij} \label{eq39b} \\ 
&&\{ \pi^i,\pi^j \} = -\frac{eb}{c} \epsilon^{ij} \label{eq39c} 
\eea
Here $\bpi$ is the kinematical (non-canonical) momentum,
$m \dot{{\bf r}}$, related to the canonical momentum $\bf p$
by ${\bpi} = {\bf p} - \frac{e}{c}{\bf A}$.

We wish to set $m$ to zero in (\ref{eq38}).
This can only be done provided $\bpi$ vanishes, and
we impose ${\bpi} = 0$ as a constraint.  But according to
(\ref{eq39c}),  the bracket of the constraints 
$C^{ij} = -\frac{eb}{c}\epsilon^{ij}$ is non-zero.
Hence we must introduce Dirac brackets:
\bea
\{ O_1, O_2 \}_D = \{ O_1, O_2 \} 
-\{ O_1, \pi^k \}(C^{-1})^{kl} \{ \pi^l, O_2 \}. 
\label{eq310}
\eea
With (\ref{eq310}), any Dirac bracket involving $\bpi$ vanishes,
so $\bpi$ may indeed be set to zero. But the Dirac bracket of two 
coordinates is now non-vanishing.
\bea
\{ r^i, r^j \}_D = -\{r^i, \pi^k\}\frac{c}{eb}
\epsilon^{kl} \{ \pi^l, r^j \}  = \frac{c}{eb}\epsilon^{ij}
\label{eq311}
\eea
In this approach, non-commuting coordinates arise as 
Dirac brackets in a system constrained to lie in the lowest 
Landau level.

\subsection{Field non-commutativity in the lowest Landau level}

We now turn to the equations of a charged fluid with density 
$\rho$ and mass parameter $m$ (introduced
for dimensional reasons) moving on a plane with velocity $\bf v$ in an
external magnetic field perpendicular to the plane. $\rho$ and
$\bf v$ are
functions of $t$ and $\bf r$ and give an Eulerian description
of the fluid.  The equations that are satisfied are the
continuity equation 
\bea
\dot{\rho} + {\bf\nabla} \cdot (\rho{\bf v}) = 0
\label{eq312}
\eea
and the Euler equation.
\bea
m \dot{v}^i + m{\bf v} \cdot {\bf\nabla} v^i = 
\frac{e}{c}\epsilon^{ij}v^jb + f^i
\label{eq313}
\eea
Here $f^i$ describes additional forces, e.g. 
$-\frac{1}{\rho} {\bf\nabla} P$ where $P$ is pressure.
We shall take the force to be derived from a potential of the 
form 
\bea
{\bf f}({\bf r}) = -{\bf\nabla} 
\frac{\delta}{\delta\rho({\bf r})} \int d^2 r V.
\label{eq314}
\eea
[For isentropic systems, the pressure is only a function of
$\rho$; (\ref{eq314}) holds with V a function of $\rho$,
related to the pressure by $P(\rho) = \rho V^{\prime}(\rho)- V(\rho)$.
Here we allow more general dependence of $V$ on $\rho$
(e.g. nonlocality or dependence on derivatives of $\rho$)
and also translation non-invariant, explicit dependence on $\bf r$.]

Equations (\ref{eq312}) and (\ref{eq313}) follow by bracketing 
$\rho$ and $\bf v$ with the Hamiltonian
\bea
H= \int d^2r \left( \rho \frac{ {\bf\pi}^2 }{2m} + V \right)
\label{eq315}
\eea
provided that fundamental brackets are taken as
\bea
&&\{ \rho({\bf r}), \rho({\bf r}^{\prime}) \} = 0 \label{eq316a} \\
&&\{ \pi({\bf r}), \rho({\bf r}^{\prime}) \} = 
{\bf\nabla} \delta({\bf r} - {\bf r}^{\prime}) \label{eq316b} \\
&&\{ \pi^i({\bf r}), \pi^j({\bf r}^{\prime}) \} =
-\epsilon^{ij}\frac{1}{\rho}\left(m \omega({\bf r}) + 
\frac{eb}{c}\right) \delta({\bf r}-{\bf r}^{\prime}) \label{eq316c}
\eea
where $\epsilon^{ij}\omega({\bf r})$ is the vorticity 
$\partial_iv^j -\partial_jv^i$, and ${\bpi} = m {\bf v}$ \cite{9}.

We now consider a strong magnetic field and take the limit 
$m\rightarrow 0$, which is equivalent to large $b$. 
Equations (\ref{eq313}) and (\ref{eq314})
reduce to
\bea
v^i = -\frac{c}{eb}\epsilon^{ij} \frac{\partial}{\partial r^j}
\frac{\delta}{\delta\rho({\bf r})} \int d^2r V.
\label{eq317}
\eea
Combining this with the continuity equation (\ref{eq312})
gives the equation for the density ``in the lowest Landau level.''
\bea
\dot{\rho}({\bf r}) = \frac {c}{eb}\frac{\partial}{\partial r^i}
\rho({\bf r}) \epsilon^{ij}\frac{\partial}{\partial r^j}
\frac{\delta}{\delta \rho({\bf r})}\int d^2r V
\label{eq318}
\eea
(For the right hand side not to vanish, V must not be solely a 
function of $\rho$.)

The equation of motion (\ref{eq318}) can be obtained by 
bracketing with the Hamiltonian
\bea
H_0 = \int d^2r V
\label{eq319}
\eea
provided the charge density bracket is non-vanishing, showing 
non-commutativity of the $\rho$'s.
\bea
\{ \rho({\bf r}), \rho({\bf r}^{\prime}) \} = 
-\frac{c}{eb}\epsilon^{ij}\partial_i\rho({\bf r})
\partial_j\delta( {\bf r}-{\bf r}^{\prime} )
\label{eq320}
\eea

$H_0$ and this bracket may be obtained from (\ref{eq315}) and
(\ref{eq316a}) -- (\ref{eq316c}) with the same Dirac procedure presented for
the particle case: We wish to set $m$ to zero in (\ref{eq315});
this is possible only if $\bpi$ is constrained to vanish.
But the bracket of the $\bpi$'s is non-vanishing, even at $m=0$,
because $b \ne 0$. Thus at $m=0$ we posit the Dirac brackets
\bea
&&\{ O_1({\bf r}_1), O_2({\bf r}_2) \}_D = \nonumber \\ 
&&\{ O_1({\bf r}_1), O_2({\bf r}_2) \} 
-\int d^2{\bf r}_1^{\prime} d^2{\bf r}_2^{\prime}
\{ O_1({\bf r}_1), \pi^i({\bf r}_1^{\prime}) \}
(C^{-1})^{ij}({\bf r}_1^{\prime}, {\bf r}_2^{\prime}) 
\{ \pi^j({\bf r}_2^{\prime}), O_2({\bf r}_2) \}
\nonumber \\
\label{eq321}
\eea
where 
\bea
(C^{-1})^{ij}({\bf r}_1, {\bf r}_2) = \frac{c}{eb}
\epsilon^{ij}\rho({\bf r}_1) \delta({\bf r}_1- {\bf r}_2). 
\label{eq322}
\eea
Hence Dirac brackets with $\bpi$ vanish, and the Dirac 
bracket of densities is non-vanishing as in (\ref{eq320}).
\bea
&&\{ \rho({\bf r}),\rho({\bf r}^{\prime}) \}_D = \nonumber \\
&&- \frac{c}{eb}\int d^2 r^{\prime\prime}
\{ \rho( {\bf r} ), \pi^i({\bf r}^{\prime\prime} \}
\rho({\bf r^{\prime\prime}}) \epsilon^{ij} 
\{ \pi^j({\bf r}^{\prime\prime}), \rho({\bf r}^{\prime}) \}= \nonumber \\
&&- \frac{c}{eb} \epsilon^{ij}\partial_i \rho({\bf r})
\partial_j\delta({\bf r} - {\bf r}^{\prime})
\label{eq323}
\eea

The $\rho$--bracket enjoys a more appealing expression in momentum space. 
Upon defining 
\bea
{\tilde{\rho}}({\bf p}) = \int d^2r e^{i{\bf p}\cdot{\bf r}}
\rho({\bf r})
\label{eq324}
\eea
we find
\bea
\{ {\tilde{\rho}}({\bf p}), {\tilde{\rho}}({\bf q}) \}_D=
-\frac{c}{eb}\epsilon^{ij}p^iq^j
\tilde{\rho}({\bf p} + {\bf q}).
\label{eq325}
\eea
The brackets (\ref{eq320}), (\ref{eq325}) give the algebra of area
preserving diffeomorphisms \cite{10}

A Lagrangian derivation, analogous to the particle case 
(\ref{eq36}) - (\ref{eq37}) is problematic
and not available.  The difficulty is that the
Poisson structures (\ref{eq316a}) -- (\ref{eq316c}) and 
(\ref{eq320}),(\ref{eq323})
are irregular: there exist ``Casimirs'' whose brackets with the 
dynamical variables vanish.  For (\ref{eq316a}) -- (\ref{eq316c})  
the Casimirs  comprise the tower 
$C^n = \int d^2 r \rho^{1-n}( m\omega+\frac{eb}{c})^n$
with $n$ arbitrary;
while for (\ref{eq320}), (\ref{eq323}) they read
$C_0^n = \int d^2r \rho^n$, again with arbitrary $n$.
(Evidently $C_0^n$ is equivalent to the $m=0$ limit of $C^n$.)
Consequently the Poisson structures do not posess an inverse; no 
symplectic 2-form can be found in terms of the above variables,
and no canonical 1-form can be added to the Hamiltonian for a 
construction of a Lagrangian. (By introducing different,
redundant variables one can remove this obstacle, at least in the
finite $m$ case) \cite{11}.

The form of the charge density bracket (\ref{eq320}), (\ref{eq323}), 
(\ref{eq325}) can be understood by reference to the particle 
substructure for the fluid. Take
\bea
\rho({\bf r}) = \sum_n \delta({\bf r} - {\bf r}_n) 
\label{eq326}
\eea
where $n$ labels the individual particles.  The coordinates of 
each particle satisfy the non-vanishing bracket (\ref{eq34}).
Then the $\{ \rho({\bf r}), \rho({\bf r}^{\prime}) \}$ bracket
takes the form (\ref{eq320}), (\ref{eq323}), (\ref{eq325}).

\subsection{Quantization of non-commutative MHD}
\label{quantize}

Quantization before the reduction to the lowest Landau level is
straightforward. For the particle case (\ref{eq39a}) -- (\ref{eq39c}) 
and for the 
fluid case (\ref{eq316a}) -- (\ref{eq316c})
we replace brackets with $i/\hbar$
times commutators.  After reduction to the lowest Landau level 
we do the same for the particle case thereby arriving at the 
``Peierls substitution,'' which states that the effect of 
an impurity [$V$ in (\ref{eq36})] on the lowest Landau energy
level can be evaluated to lowest order by viewing the $(x,y)$ arguments 
of $V$ as non-commuting variables \cite{2}.

However, for the fluid case quantization presents a choice.
On the one hand, we can simply promote the 
bracket (\ref{eq320}), (\ref{eq323}), (\ref{eq325}) to a 
commutator by multiplying by $i/\hbar$.
\bea
&&[ \rho({\bf r}),\rho({\bf r}^{\prime}) ] = 
i\hbar\frac{c}{eb} \epsilon^{ij}\partial_i \rho({\bf r}^{\prime})
\partial_j\delta({\bf r} - {\bf r}^{\prime}) \label{eq327a} \\
&&\left[ \tilde{\rho}({\bf p}), \tilde\rho({\bf q}) \right] =
i \hbar \frac{c}{eb} \epsilon^{ij} p^i q^j \tilde\rho({\bf p} + {\bf q})
\label{eq327b}
\eea

Alternatively we can adopt the expression (\ref{eq326}), for the operator 
$\rho({\bf r})$, where the ${\bf r}_n$ now satisfy the
non-commutative algebra
\bea
\left[  r_n^i,  r_{n^{\prime}}^j \right] = -i\hbar \frac{c}{eb} 
\epsilon^{ij}\delta_{nn^{\prime}}
\label{eq328}
\eea
and calculate the $ \rho$ commutator as a derived
quantity.

However, once ${\bf r}_n$ is a non-commuting operator, functions
of $ {\bf r}_n$, even $\delta-$functions,
have to be ordered. We choose the Weyl ordering,  which is equivalent to 
defining the Fourier transform as 
\bea
\tilde{\rho}({\bf p}) = \sum_n e^{i{\bf p} \cdot {\bf r}_n}.
\label{eq329}
\eea
With the help of (\ref{eq328}) and the Baker-Hausdorff lemma, we
arrive at the ``trigonometric algebra'' \cite{12}
\bea
[\tilde{\rho}({\bf p}), \tilde{\rho}({\bf q})]=
2i \sin \left( \frac{\hbar c}{2eb} \epsilon^{ij}p^iq^j \right)
\tilde{\rho}({\bf p}+ {\bf q}).
\label{eq330}
\eea
This reduces to (\ref{eq327b}) for small $\hbar$.

This form for the  commutator, (\ref{eq330}), 
is connected to a Moyal star product \cite{13} in
the following fashion.  For an arbitrary c-number function $f({\bf r})$
define
\bea
<f> = \int d^2r \rho({\bf r})f({\bf r}) = 
\frac{1}{(2\pi)^2}\int d^2 p \tilde{\rho}({\bf p}) \tilde{f}(-{\bf p}).
\label{eq331}
\eea
Multiplying (\ref{eq330}) by 
$\tilde{f}(-{\bf p})\tilde{g}(-{\bf q})$ and integrating
gives
\bea
[<f>, <g>] = <h>,
\label{eq332}
\eea
with
\bea
h({\bf r}) = (f*g)({\bf r}) - (g*f)({\bf r})
\label{eq333}
\eea
where the ``$*$'' product is defined as 
\bea
(f*g)({\bf r}) = e^{\frac{i}{2}\frac{\hbar c}{eb} 
\epsilon^{ij} \partial_i \partial_j^{\prime}}
f({\bf r})g({\bf r}^{\prime})|_{ {\bf r}^{\prime} = {\bf r}}.
\label{eq334}
\eea
Note however that only the commutator is mapped into the star 
commutator.  The product $<f><g>$ is not equal to $<f*g>$.

The lack of consilience between (\ref{eq327b}) and (\ref{eq330})
is an instance of the Groenwald-VanHove theorem which establishes 
the impossibility of taking over into quantum mechanics all classical
brackets \cite{13}. Equations (\ref{eq330}) -- (\ref{eq334}) explicitly
exhibit the physical occurence of the star product for fields in
a strong magnetic background.

\section*{Acknowledgments}

Z.G. thanks W. Skiba for discussions.
R.J. acknowledges S. Carroll for information on ref.\cite{5}.
A.P. is grateful to City College, CUNY, for hospitality during
part of this work.  This work is supported in part
by funds provided by the U.S. Department of Energy (D.O.E.) under
cooperative research agreements DE-FC02-94ER40818 and
DE-FG02-91ER40676.


\end{document}